# Digital Twin for Grey Box modeling of Multistory residential building thermal dynamics

Lina Morkunaite
Faculty of Civil Engineering and Architecture
Kaunas University of Technology
Kaunas, Lithuania
lina.morkunaite@ktu.lt

Justas Kardoka
Faculty of Informatics
Kaunas University of Technology
Kaunas, Lithuania
justas.kardoka@ktu.lt

Darius Pupeikis
Faculty of Civil Engineering and Architecture
Kaunas University of Technology
Kaunas, Lithuania
darius.pupeikis@ktu.lt

Paris Fokaides
School of Engineering
Frederick University
Nicosia, Cyprus
p.fokaides@frederick.ac.cy

Vangelis Angelakis
Department of Science and Technology
Linkoping University
Norrkoping, Sweden
vangelis.angelakis@liu.se

*Abstract*—Buildings' energy efficiency is a widely researched topic, which is rapidly gaining popularity due to rising environmental concerns and the need for energy independence. In Northern Europe heating energy alone accounts for up to 70% of the total building energy consumption. Industry 4.0 technologies such as IoT, big data, cloud computing and machine learning, along with the creation of predictive and proactive digital twins, can help to reduce this number. However, buildings' thermal dynamics is a very complex process that depends on many variables. As a result, commonly used physics-based white box models are time-consuming and require vast expertise. On the contrary, black box forecasting models, which rely primarily on building energy consumption data, lack fundamental insights and hinder re-use. In this study we device an architecture to facilitate grey box modeling of building's thermal dynamics while integrating real time IoT data with 3D representation of buildings. The architecture is validated in a case study creating a digital twin platform that enables users to define the thermal dynamics of buildings based on physical laws and real data, thus facilitating informed decision making for the best heating energy optimization strategy. Also, the created user interface enables stakeholders such as facility managers, energy providers or governing bodies to analyze, compare and evaluate buildings' thermal dynamics without extensive expertise or time resources.

I. INTRODUCTION

Buildings as one of the main energy consumers within the power grid are at the forefront of the energy optimization strategies in the EU [1]. Heating represents the largest share of energy consumption in buildings, which is a major environmental concern, as a large share of heat is still produced by using fossil fuels [2]. Various solutions are proposed and adopted to address this problem, such as buildings' envelopes retrofit, energy control interventions, integration of renewable energy sources, and so forth [3], [4]. However, buildings are complex systems whose performance highly depend on thermal dynamics. Therefore, adapting an appropriate energy optimization strategy becomes a complex and tedious task, which often leads to decisions based on approximate rather than real performance data [5].

Opposed to the commonly used white box models for heat dynamics, which are based on physical laws of heat transfer as conduction, convection and radiation, data-driven black box models have recently gained considerable popularity [6]–[8]. Though black box models have shown great results in the analysis of large amounts of data, the main limitations still lies in their interpretation, validation and reusability [9]. To reduce the time and expert resources required for white box models and increase the understanding of physical laws within the black box models, a grey box modelling approach was introduced [10]. Grey box modelling combines physical laws and real-life data making the best use of both former methods. A recent study of 247 residential buildings in the Netherlands showed that grey box modelling can be used to automatically select the most appropriate thermal dynamics model for a building [5]. Even though not all buildings showed a good model fit, this study has opened the door to many future applications for heating energy optimization strategies.

Nevertheless, to make this approach widely available to stakeholders such as facility managers, energy providers or governing bodies, it is necessary to deliver a solution enabling data integration and access to model fitting. Regardless the immense developments brought by Industry 4.0 for the built environment in terms of big data management, IoT and other technologies, data integration still remains a challenge [11], [12]. Combining the 3D representation of buildings with real-time data results in a digital twin, which enhanced with heat dynamics modelling enables proactive decision making. Such digital twins can be applied for various strategies to optimize heating energy, to evaluate buildings' energy performance in comparison to its peers or manage energy in crisis conditions. In fact, a need for such remote tools was clearly seen during the recent pandemic to ensure smooth energy supply, maintain energy balance or deal with system failures.

This study presents a real time data integration architecture for time series data of heating energy consumption, indoor air temperature, outdoor ambient air temperature and solar irradiation, to be used for grey box modelling of building's thermal dynamics. To enhance the decision making process, the dataset is supplemented with a reality capture 3D representation of the case study buildings' complex, as well as a user interface (UI) to visualize the data in tables and interactive graphs of the individual datasets as buildings' digital twin. To enable the integration several APIs are developed, while the visualization and UI are facilitated by *Bentley's iTwin* platform. The proposed architecture is further tested and validated in a pilot case which is a part of Horizon2020 project "Development of Utilities Management Platform for the case of Quarantine and Lockdown (eUMaP)".

## II. METHODOLOGY

### A. Data integration

The proposed architecture for data integration and analysis is combined from several layers including the physical assets, data sources, integration layer, data storage layer, simulation layer and UI (Fig. 1). To integrate data from IoT devices (sensors, energy meters, etc.) and other data with buildings' photogrammetry 3D model, it is worth considering the use of highly customizable platforms that provide tools for developing distinct features. In this work, *Bentley's iTwin* platform was employed for its extensive functionality while developing digital twins. *iTwin* supports various 3D representation file formats (e.g., DGN, IFC, DWG) and provides open APIs for additional data incorporation to the model. The *iTwin* solution is developed using *HTML, CSS, JavaScript, TypeScript* programming languages and *React* library that can later be deployed as a web application. The users can openly access and exploit the features of the web application, however, to make changes a license is required.

The integration of 3D geometrical representation model of buildings into *iTwin* was carried out through the exchange of different file formats. First, a DGN file photogrammetry mesh model was generated using *Context Capture* software that is native to *Bentley Systems*. Afterwards the conversion into *iModel* through the *iTwin* synchronizer tools was needed. Finally, the prepared model was stored in *iModelHUB*, which is a cloud service enabling alignment, accountability and accessibility of infrastructure digital twins and made available for further development using *iTwin* platform.

To emulate real-time data for heating energy consumption from the energy provider's smart meters, a time period of 1 year was extracted from the dataset with frequency of 1 data point per hour. To match the heating energy consumption data to 15 min data points collected by indoor air temperature sensors, the hourly energy consumption values were equally distributed for 4 data points through developed API. Indoor air temperature data were collected by the physically installed sensors and through the Oracle server broadcasted into *Zabbix* middleware in JSON format. Another API was used to retrieve the weather data of outside air temperature and solar irradiation. The data were collected from an openly available *Oikolab* online repository and broadcasted into *Zabbix* platform.

Parameter estimation and model fitting of the grey box building heat dynamics models was done using the R programming language in *RStudio* environment adapting the solution from [5]. To integrate R models into *iTwin* platform a dedicated API was developed using an open source *Swagger* toolset and *Plumber* API generator for R. The developed API also allows the user to access historical and real-time heating energy consumption, indoor air temperature and outdoor climate conditions data from *Zabbix* middleware. The data is further used to fit a grey box model for a selected building, while extracting user defined time period via developed UI.

### B. Grey box modelling

Grey box models for heat dynamics allow the estimation of buildings thermal capacity, resistance and useful window area in continuous time while using discrete time data points. The models are built by applying the physical laws of thermal conduction, convection and radiation while adding the concept of noise to account for the residuals. They are written in the form of stochastic differential equations (SDEs), that allows adjusting the model complexity to find the most suitable model defining the heat dynamics of a building.

Grey box models are developed based on RC (resistance-capacitance) lumped representation of the system and are combined from system and measurements equations. P. Bacher and H. Madsen suggest five model complexities, predicting indoor, medium, heater, envelope and sensor temperatures [10]. An expression for the first-order stochastic differential equation is as follows:

$$dT_i = \frac{1}{R_{ia}C_i}(T_a - T_i)dt + \frac{1}{C_i}\Phi_h dt + \frac{1}{C_i}A_w\Phi_s dt + \sigma_i dw_i \quad (1)$$

Where the T is the temperature, R the resistance and C the capacitance. $\Phi$ is the energy flux and $A_w$ is the useful window area. The $i$ in the subscript represents the indoor air, *a* the ambient air, *h* the heater and *s* the sensor temperatures. The standard Wiener process $w$ and incremental variances of the Wiener process $\sigma$ are described more in detail by [10].

Equation (1) (also called the system equation) defines the deterministic part of the model. This part is supplemented by the stochastic part called the measurement equation:

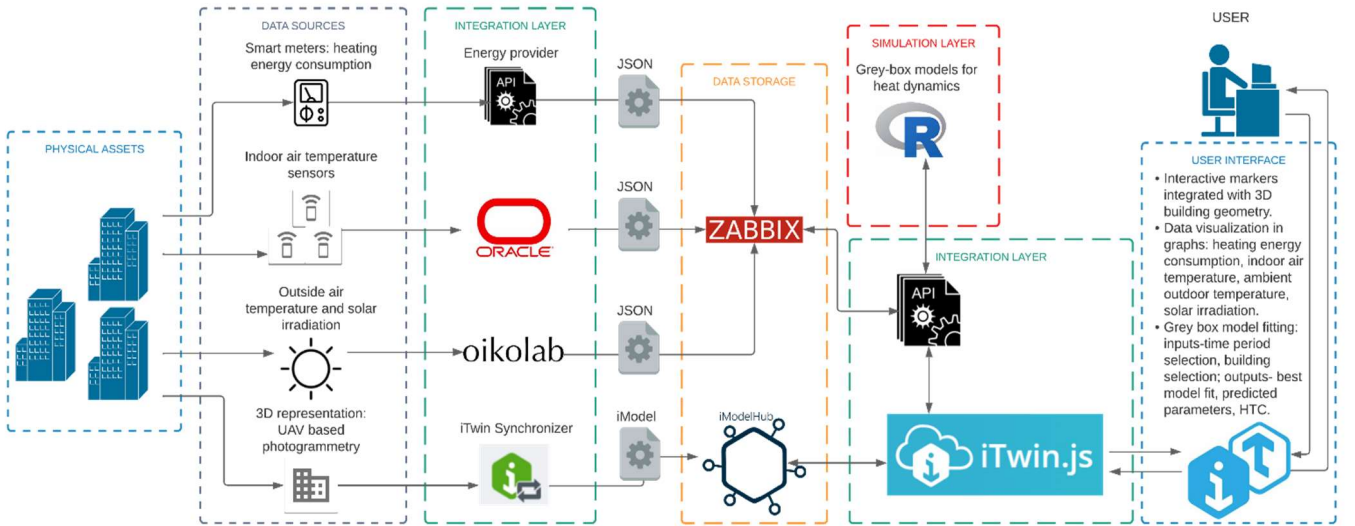

Fig. 1. Data integration architecture.

$$Y_t = T_{i,t} + e_t \quad (2)$$

Where $Y_t$ is the measured indoor air temperature, $t$ the measurement point in time and $e_t$ the measurement error, which is assumed to be Gaussian white noise [10].

The model fitting follows a forward selection process where based on likelihood ratio tests it is determined if the simpler model performs better than the more complex one. The optimization process is terminated when no model extension yields a higher p-value than 5% [5].

The full model $T_iT_mT_eT_hT_sA_eR_{ia}$ consists of the five SDEs where *i, m, e, h, s* are the temperatures of indoor, medium, envelope, heater and sensor, while $A_eR_{ia}$ account for the solar gains. The five SDEs used in this study are presented by [5]:

$$dT_s = \frac{1}{R_{is}C_s}(T_i - T_s)dt + \sigma_s dw_s \quad (3)$$

$$dT_i = \frac{1}{R_{is}C_i}(T_s - T_i)dt + \frac{1}{R_{im}C_i}(T_m - T_i)dt + \frac{1}{R_{ih}C_i}(T_e - T_i)dt + \frac{1}{R_{ie}C_i}(T_e - T_i)dt + \frac{1}{R_{ia}C_i}(T_a - T_i)dt + \frac{1}{C_i}A_w\Phi_s dt + \sigma_i dw_i \quad (4)$$

$$dT_m = \frac{1}{R_{im}C_m}(T_i - T_m)dt + \sigma_m dw_m \quad (5)$$

$$dT_h = \frac{1}{R_{ih}C_h}(T_i - T_h)dt + \sigma_h dw_h \quad (6)$$

$$dT_e = \frac{1}{R_{ie}C_e}(T_i - T_e)dt + \frac{1}{R_{ea}C_e}(T_a - T_e)dt + \frac{1}{C_e}A_e\Phi_s dt + \sigma_e dw_e \quad (7)$$

The measurement equation:

$$Y_t = T_{s,t} + e_t \quad (8)$$

The model optimization is run in *RStudio* and via constructed API the estimated building thermal capacitance, resistance and useful window area is sent back to the user. During the process the heat transfer coefficient (HTC) is estimated as well. The user can validate if the model was well fit by checking if the residuals are independent and identically distributed, so called white noise. This can be inspected in the measured/predicted indoor temperature with other parameters comparison plots, as well as in the cumulated periodogram, boundary excess and cumulated nCPBES (cumulated periodogram boundary excess sum) graphs provided by the UI. The nCPBES parameter was proposed to estimate the goodness of the model fit and is introduced in detail by [5].

### C. UI development in iTwin

The user interface was constructed using *iTwin.js* library, which is a set of APIs, that allows to aggregate, visualize and analyze data in *iTwin* environment. Separate interactive windows were created to view the datasets for heating energy consumption, indoor air temperature, ambient outdoor air temperature and solar irradiation in tabular and graph formats. Data graphs can be inspected by turning off and on the required data types, zooming in/out or using the *pan* function. If needed the inspected graph can be downloaded in a PNG format and the initial graph view can be reset by pressing the home button. A marker was created on top of the building's 3D photogrammetry model that by hoovering the cursor on top of it displays the latest sensor data for indoor air temperature, humidity and volatile organic compounds (VOC). By clicking the marker, the user can inspect all 3 indoor air quality (IAQ) parameters in separate graph views.

All the datasets for different buildings were listed on the right-hand side of UI. The datasets can be refreshed by using the *Refresh data* button, which also shows the time stamp for when the data was refreshed last. To fit a grey box model for heat dynamics another interactive window was created, that opens up while clicking on the required building *Forecasts* button. The pop-up window allows the user to choose a time period from the calendar view to be extracted from the dataset that will be used for heat dynamics model fitting.

When the optimization is complete an additional window was created which returns the user the best model fit name with predicted parameters of resistances, capacitances and useful window area as well as calculated heat transfer coefficient (HTC). To evaluate the goodness of model fit the user can inspect the plots of cumulated periodogram, boundary excess and cumulated nCPBES parameter which was sourced from the R code. Additionally, predicted parameters for residuals and measurement error can be reviewed. One more graph was created to inspect the correlation coefficients between the predicted parameters. Finally, the plots of actual and predicted indoor air temperature were incorporated to analyze the precision of predictions, that can also be evaluated together with plots for the input parameters of thermal energy consumption, outdoor air temperature and solar irradiation.

### III. CASE STUDY

#### A. Case study building

The proposed data integration and heat dynamics grey box model evaluation architecture was tested on a multi-story residential apartment building that is a part of a modern buildings complex in Kaunas, Lithuania (Fig. 2).

The building shell is EPS insulated masonry wall structure, plastic frame windows with double glazing unit, mineral wool insulated flat roof structure and ground floor structure of reinforced concrete with vertical and horizontal XPS insulation. The specific values of the thermal transmittances and materiality of the building enclosures are given in Table 1. All the buildings in the complex are connected to the district heating system (DHS) and has their own heat distribution point.

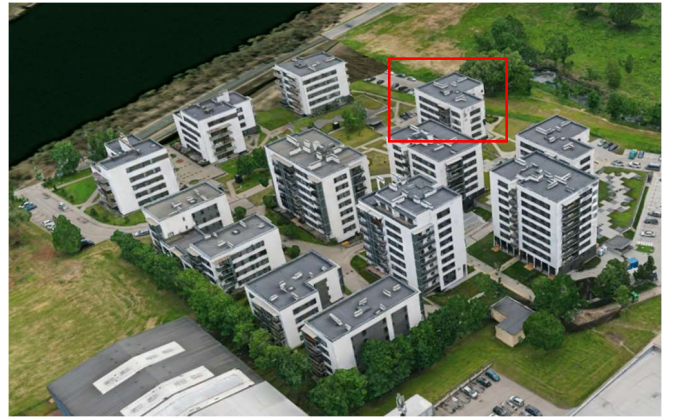

Fig. 2. Case study apartment buildings' complex photogrammetry model (investigated building marked by a red rectangle).

TABLE I. THERMAL TRANSMITTANCE VALUES AND MATERIALITY OF THE BUILDING ENCLOSURES.

| Element | Thermal transmittance, W/(m²·K) | Materiality |
|---|---|---|
| Walls | 0.2 | Ceramic block masonry, EPS thermal insulation, plaster. |
| Windows | 1.1 | Plastic frame with double glazing unit filled with argon gas. |
| Roof | 0.16 | Reinforced concrete slabs, expanded clay layer, EPS thermal insulation, wind-proof rock-wool insulation, PVC hydro insulation. |
| Ground floor | 0.25 | Reinforced concrete, EPS thermal insulation, PVC indoor coating. |

### B. Data collection

The building's heating energy consumption is retrieved from the local energy provider, while the data regarding indoor air temperature is collected via installed sensors. Buildings' geometry is represented using a 3D photogrammetry model and the weather condition data (ambient outdoor air temperature and solar irradiation) is collected from *Oikolab* online repository.

## IV. RESULTS AND DISCUSSION

### A. Data visualization in iTwin platform

The data integration architecture and developed UI was tested using *iTwin* platform for the case study building. The application was accessed via local host on a web browser. The *iTwin* environment provides the user with traditional 3D viewing functionalities to inspect the photogrammetry model as rotation, pan and zoom. Geometric representation was useful to get a general understanding of the building shell (basic structure, wall/window ratio, etc.) and its surrounding elements. Additionally, it was possible to use the measurement tool to approximately estimate the building size and glazed area. However, no semantic information for the 3D photogrammetry model was available.

Indoor air quality sensor data was accessed by selecting the developed marker on the imported 3D photogrammetry model of the building under analysis. As the IAQ sensor collects data not only on indoor air temperature but also on indoor air humidity and air quality (through volatile organic compounds measurements), these values and their trends in the graph format were also inspected (Fig.3). Similarly, indoor air temperature was verified in a table format next to the heating energy consumption, outside air temperature and solar irradiation data in tabular and graph formats (Fig. 4 and 5). In the measured data plots for the afore mentioned parameters (Fig. 5) dependencies between the outdoor air temperature and heating energy consumption was clearly observed.

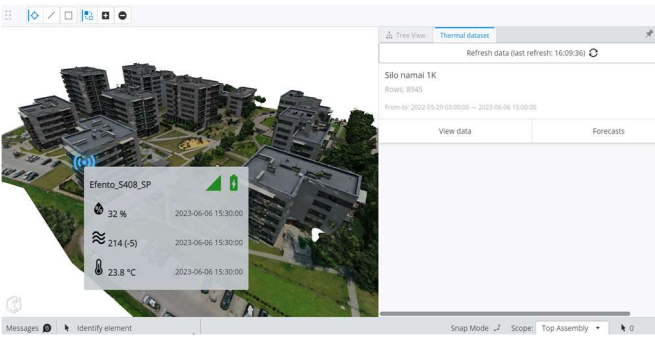

Fig. 3. 3D photogrammetry model with sensor marker providing data on indoor air quality within *iTwin* environment.

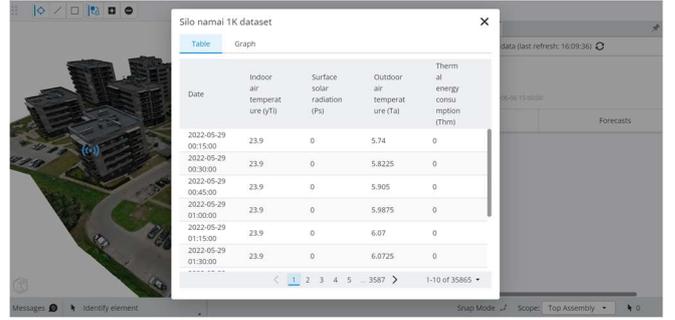

Fig. 4. Analyzed dataset in table format within *iTwin* environment.

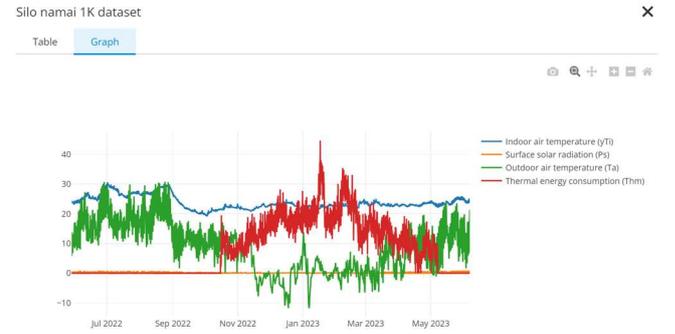

Fig. 5. Input parameters data plots in *iTwin* environment.

At the same time, the diagrams allow some assumptions to be made about thermal inertia and capacity of the building. It is visible that heating of the building starts before outdoor air temperature reaches very low temperatures, which is common to preheat the building shell. Also, there is a noticeable lag when temperatures drop and heating energy increases.

### B. Grey box model for building heat dynamics

Unlike the black box modelling, grey box models can be fitted with fewer amount of data [10]. To reduce the requirements for computational resources while running optimization only one week of data was extracted from the dataset ranging from the 8th to 15th January 2023. The data range was set using the *Forecasts* button and by selecting the required dates in the calendar provided by UI.

The final input dataset consisted of 764 datapoints every 15 min for the indoor temperature, outdoor ambient temperature, heating energy consumption and solar irradiation. The model fitting followed a forward selection process starting with the simplest model and increasing the complexity until no more significant dependencies could be found within the residuals. The statistical methods were facilitated by CTSMR (continuous time stochastic modeling in R) library developed by Danish Technical University [13].

For the selected building dataset, it took 1 minute and 22 seconds to fit the model, but it was observed that a larger data set significantly increases the model fitting time. At the same time, if several buildings are run simultaneously during the optimization, the model fitting time is also prolonged. This was found to be one of the main drawbacks for this solution. However, it could be overcome by using a parallel loop rather than in-series loop as suggested by [5].

Another potential issue could be raised by the iterative process of model selection which is bounded by initial parameter declaration embedded within the code. The declaration defines 5 initial values for each of the estimated

```
# Initial parameter declaration
inl <- c()
inl[["Tm_ini"]] <- c(19, 19, 19, 20, 18)
inl[["Tm_lb"]]  <- c(12, 12, 10, 5, 5)
inl[["Tm_ub"]]  <- c(30, 30, 30, 35, 40)
inl[["Te_lb"]]  <- c(-10, -10, 0, 0, 0)
inl[["Te_ub"]]  <- c(40, 40, 40, 35, 35)
inl[["Ci_ini"]] <- c(5, 1, 3, 10, 15, 50)
inl[["Ci_ub"]]  <- c(20, 30, 40, 60, 100, 300)
inl[["Ria_ini"]] <- c(5, 5, 5, 10, 1)
inl[["Rie_ini"]] <- c(5, 10, 10, 10, 1)
inl[["Rea_ini"]] <- c(5, 5, 10, 10, 5)
inl[["Rih_ini"]] <- c(5, 5, 10, 10, 1)
inl[["Rim_ini"]] <- c(1, 1, 3, 5, 1)
inl[["Ris_ini"]] <- c(1, 1, 3, 5, 1)
inl[["Rh_ini"]]  <- c(0.5, 0.5, 0.5, 1, 2)
inl[["p_ini"]]   <- c(1, 1, 1, -1, 5)
inl[["p_lb"]]    <- c(-40, -40, -40, -50, -20)
inl[["p_ub"]]    <- c(20, 20, 20, 10, 40)
inl[["e_ini"]]   <- c(-1, -1, 1, 1, 0.5)
inl[["e_lb"]]    <- c(-50, -40, -40, -40, -20)
inl[["e_ub"]]    <- c(20, 20, 20, 30, 40)
```

Fig. 6. Boundary conditions for parameters set in RStudio.

TABLE II. FITTED MODEL PARAMETER VALUES.

| Parameter | Parameter meaning | Result | Unit |
|---|---|---|---|
| Ti | Indoor temperature | 22.60 | °C |
| Th | Heater temperature | 23.19 | °C |
| Tm | Heating medium temperature | 23.27 | °C |
| Ci | Indoor capacitance | 0.76 | kWh/K |
| Ch | Heater capacitance | 9.51 | kWh/K |
| Cm | Heating medium capacitance | 19.86 | kWh/K |
| Rih | Heater resistance | 0.02 | K/kW |
| Ria | Window resistance | 49.66 | K/kW |
| Rim | Heating medium resistance | 2.53 | K/kW |
| Aw | Useful window area | 2.83 | m² |
| HTC | Heat transfer coefficient | 0.02 | kW/K |

parameters including the lower and upper boundaries (Fig. 6). For example, the highest value for thermal capacitance is set to be 300 kWh/K. Even though, this declaration was employed evaluating a variety of buildings from single-family houses to multi-story buildings in the Dutch case study [5], in practice the upper boundary for large buildings thermal capacity can span outside the proposed value. In such case CTSMR package displays a warning in *RStudio* environment: *"The upper boundary has been hit!"*. However, at the current state of proposed solution, the warning is not visible to the user.

The grey box model fitting is based on likelihood ratio test which allows to determine if the former model statistically performs better than the more complex one. When the more complex model does not demonstrate a *p* value larger than 5% the improvement is considered statistically insignificant, and the former model is saved as the best fit. In this case the process was terminated under the 3rd iteration and the suitable model was selected as *TiThTm*. The selected model parameter values are shown in Table 2.

Further, the model fit was evaluated via statistical methods. Ranging from the first to the last iteration the nCPBES parameter never showed a greater value than 0.01. At the last iteration the nCPBES parameter showed a value of 0 which classifies the model as a good fit. According to [5] the closer the nCPBES parameter comes to 0 the better is the fit. A good model fit is classified as one below 0.01, medium fit is below 0.03 and all above are classified as a poor fit. The residual mean was determined to be 0.00096 and standard deviation equal to 0.04. The quality of the fit was also inspected in graphs. The cumulated periodogram showed that the selected model did not exceed the 95% confidence bounds, therefore no boundary excess is shown in the subsequent graph. In the plotted time-series graph it was observed that the predicted indoor air temperature *(Ti)* values very closely matched the measured values, and no significant correlations were seen with other input parameter plots. Thus, an assumption can be made that the residuals were white noise. The correlations between the parameters were also evaluated using the correlation heatmap provided by UI. The results for the estimated model fit with its parameters including the heat transfer (HTC) coefficient and the graphs for statistical methods within *iTwin* environment are presented in Fig. 7.

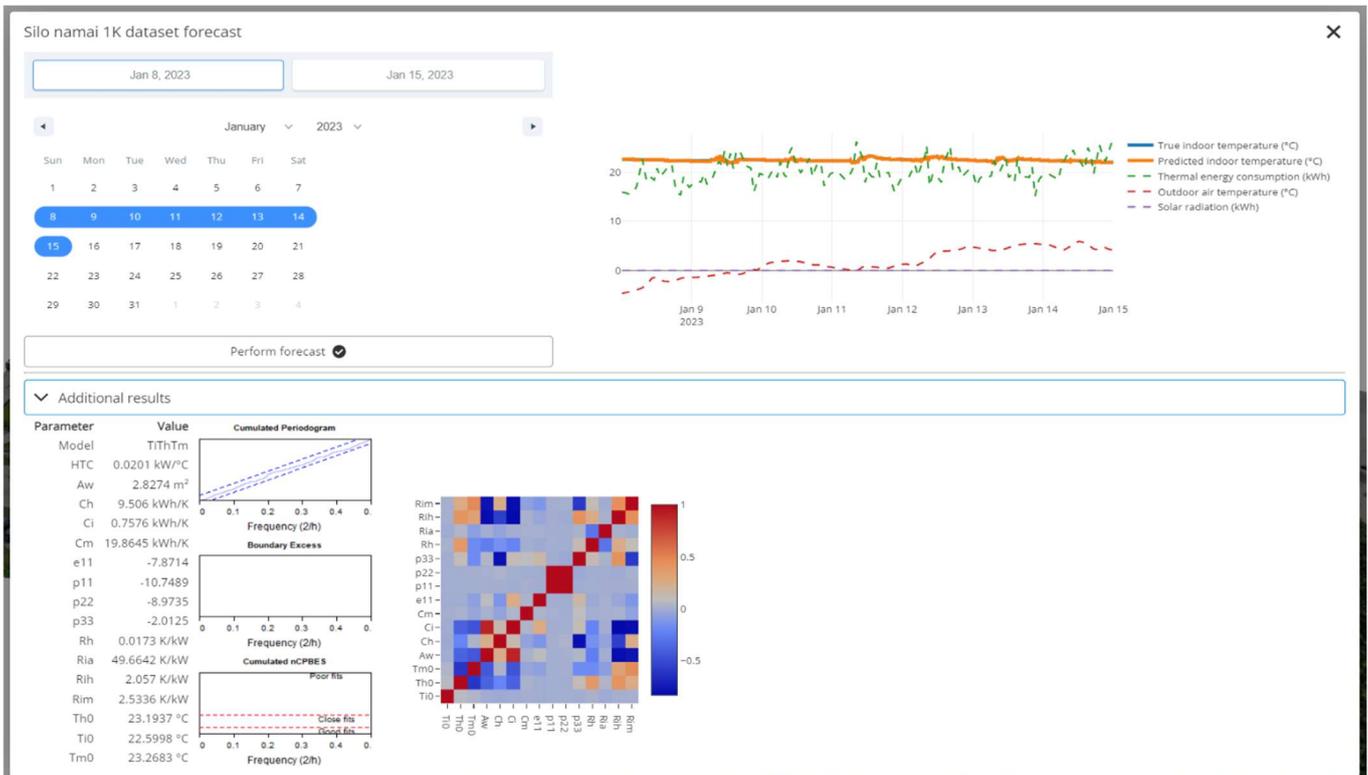

Fig. 7. Results for the best model fit, estimated parameters and plots for model evaluation by statistical methods.

## V. Conclusions

This work presented an architecture for integrating different data sources that allows to define building's thermal dynamics, such as building thermal energy consumption, indoor air temperature, outdoor air temperature and solar radiation, into a single platform by creating separate APIs. The architecture was tested through a case study where real-time sensor data and a 3D photogrammetric model of a residential block of flats were accessed and visualized through the *Bentley's iTwin* platform. The integrated time-series data were also accessed via another API allowing to fit a grey box model facilitated by *RStudio*. A user interface was developed that enables the user to check and compare different data sources presented in tables and graphs; to see a timestamp of when the data was last updated and, if necessary, to update it at the click of a button; to select the desired period of time from the dataset and to check the corresponding graphs; to fit a grey box model using the selected time period data; to check the results of the fitted model (resistances, capacitances, useful window area, and heat transfer coefficient); and to assess the model fit by analyzing it using statistical methods represented in graphs.

Even though, there are still challenges to overcome for proposed solution as the computing resources to fit larger datasets or analyze several at once; parameter lower and upper boundary conditions adjustments availability for the user; input data preparation and real-time data management; or availability of buildings' 3D photogrammetry model semantic data. The results showed that such a solution can allow the stakeholders as facility managers, engineers or governing bodies to make more informed decisions while choosing the most suitable energy optimization strategy for buildings or manage DHS during crisis conditions. Using the integrated platform users can evaluate buildings' thermal dynamics without extensive expertise in calculating white-box models and at the same time gain insights based on physical laws.

In the future more buildings' data should be integrated allowing the users to compare the model fits, heating energy consumption and indoor air temperature enabling buildings peer-to-peer analysis. Further the models could be used to predict energy consumption of the buildings allowing implementation of such strategies as renewable energy sources integration or energy management in crisis conditions.

## Nomenclature

| Symbol | Description | Unit |
|---|---|---|
| $T$ | Temperature | °C |
| $R$ | Resistance | K/kW |
| $C$ | Capacitance | kWh/K |
| $A_w$ | Effective window area | m$^2$ |
| $\Phi_h$ | Heating system energy flux | W/m$^2$ |
| $\sigma$ | Incremental Wiener process variance | - |
| $w$ | Standard Wiener process | - |
| $Y_t$ | Measured indoor temperature | °C |
| $e_t$ | Measurement error | - |
| Indicator | | |
| $i$ | Indoor | - |
| $a$ | Ambient | - |
| $e$ | Envelope | - |
| $h$ | Heater | - |
| $s$ | Sensor | - |

## Acknowledgment

This project has received funding from the SMART-ER project, funded by the European Union's Horizon 2020 research and innovation program under Grant Agreement #101016888.

The idea for the utilities data integration platform was initially generated during the project "Development of Utilities Management Platform for the case of Quarantine and Lockdown" (Grant agreement ID: 101007641), funded by the European Commission.